\documentclass[journal]{IEEEtran}
\usepackage{multirow}
\usepackage{amsmath,amsfonts}
\usepackage{algorithmic}
\usepackage{algorithm}
\usepackage{array}
\usepackage[caption=false,font=normalsize,labelfont=sf,textfont=sf]{subfig}
\usepackage{textcomp}
\usepackage{stfloats}
\usepackage{url}
\usepackage{verbatim}
\usepackage{graphicx}
\usepackage{cite}
\usepackage{adjustbox}
\usepackage{tabularx}
\usepackage{booktabs}
\usepackage{hyperref}

\usepackage{balance}
\usepackage{soul}
\usepackage{xcolor}
\begin{document}

\title{Context-Dependent Autonomic Responses in Social Anxiety During Cognitive-Emotional Stress}

 \author{
Arya Adyasha*,
Vikas Kumar*,
Anushka Sanjay Shelke,
and Haroon R. Lone%
\thanks{
* equal contribution
}
\thanks{
All authors are from the Indian Institute of Science Education and Research Bhopal, Bhopal, India.
}
\thanks{
Corresponding author: Haroon R. Lone (haroon@iiserb.ac.in).
}

}

\markboth{Journal of \LaTeX\ Class Files,~Vol.~14, No.~8, August~2021}%
{Anonymous \MakeLowercase{\textit{et al.}}: Autonomic Arousal in Social Anxiety}

\maketitle

\begin{abstract}

Social anxiety disorder (SAD) is associated with heightened physiological arousal during socially evaluative situations, yet it remains unclear whether similar autonomic responses emerge during non-evaluative cognitive-emotional stress. This study investigated wearable electrodermal activity (EDA) responses in socially anxious (SA) and non-socially anxious (NSA) individuals during an emotionally salient 2-back working memory task involving facial expressions. Fifty participants (25 SA, 25 NSA) completed a resting-state baseline and task condition while EDA signals were acquired using a Shimmer3 GSR+ sensor. EDA features spanning tonic, phasic, sympathetic, spectral, and nonlinear domains were analyzed using mixed ANOVAs and complementary machine learning models. Results showed significant increases in autonomic arousal during task engagement across all participants, confirming that the task induced substantial sympathetic activation. However, no consistent between-group differences were observed, with only transient interaction effects emerging during the initial task phase. Machine learning analysis demonstrated above-chance discrimination between SA and NSA individuals using resting-state EDA (average AUC~=~0.73), whereas classification performance during task engagement declined to near-chance levels (average AUC~$\leq$~0.57). These findings suggest that cognitively demanding emotional tasks, in the absence of explicit social-evaluative threat, elicit comparable autonomic responses regardless of social anxiety status and may obscure subtle resting-state physiological differences between groups. More broadly, our findings highlight the context-dependent nature of wearable autonomic biomarkers for anxiety assessment and digital mental health monitoring. With this manuscript, we release both the code and data publicly.
\end{abstract}

\section{Introduction}


Social anxiety disorder (SAD) is a prevalent mental health condition characterized by persistent fear of social scrutiny and negative evaluation, often leading to impaired social, cognitive, and emotional functioning~\cite{wells1995social, rapee1997cognitive}. Affecting approximately 4.7--17\% of the global population~\cite{salari2024global}, SAD substantially impacts quality of life, academic performance, interpersonal relationships, and occupational functioning~\cite{kessler2005lifetime}. In addition to psychological symptoms such as anticipatory anxiety, fear of embarrassment, and post-event rumination~\cite{stein2008social}, individuals with SAD frequently exhibit heightened autonomic nervous system (ANS) activation during socially demanding situations~\cite{perna2020personalized}. These physiological manifestations include sweating, trembling, increased heart rate, nausea, muscle tension, and dizziness. The growing need for objective and scalable approaches for anxiety assessment has therefore motivated increasing interest in wearable physiological sensing technologies for digital mental health monitoring.

Electrodermal activity (EDA), a non-invasive measure of sympathetic autonomic arousal, has emerged as a promising physiological marker for stress and anxiety assessment in wearable health systems~\cite{boucsein2012electrodermal}. EDA reflects changes in skin conductance caused by sweat gland activity regulated by the sympathetic nervous system and is commonly decomposed into tonic skin conductance level (SCL) and phasic skin conductance responses (SCRs)~\cite{sachdeva2022electrodermal}. Because EDA can be continuously acquired using wearable sensors in an unobtrusive manner, it has gained increasing attention as a potential digital biomarker for affective and mental health monitoring. Prior studies have reported associations between elevated skin conductance and anxiety-related states~\cite{mauss2003autonomic, wu2013impact, najafpour2017can}. However, findings remain inconsistent across different experimental paradigms and social contexts~\cite{jensen1996electrodermal, gungor2020effect, sahu2024wearable}, limiting the reliability and generalizability of EDA-based anxiety assessment frameworks.

Several studies have attempted to characterize autonomic differences between socially anxious and non-anxious individuals under stress-inducing conditions. For instance, Moscovitch et al. reported heightened SCR responses in individuals with elevated anxiety during public speaking tasks~\cite{moscovitch2010emotional}, whereas Eckman et al. observed no significant SCL differences between high- and low-anxiety participants during similar speech-based activities~\cite{eckman1997habituation}. A recent systematic review by Fischer et al.~\cite{fischer2021systematic} further highlighted the inconsistency of skin conductance findings in social anxiety research, particularly across experimentally induced social stress paradigms. Such variability suggests that physiological responses in socially anxious individuals may depend strongly on contextual factors, including the presence of social-evaluative threat, cognitive demand, and individual coping responses~\cite{white2018low, panayiotou2017psychophysiological, heiser2009differentiating}. Understanding these context-dependent autonomic patterns is important for the development of reliable wearable mental health monitoring systems.

Cognitive theories of social anxiety further suggest that individuals with SAD experience impairments in emotional and interpersonal information processing, which may influence attentional control and working memory performance~\cite{moran2016anxiety}. To examine anxiety-related responses under cognitively demanding conditions, prior studies have frequently employed the n-back task~\cite{braver1997parametric, smith1997working}, a working memory paradigm requiring participants to continuously monitor and update incoming stimuli. Emotionally salient versions of the n-back task, particularly those involving facial expressions~\cite{levens2010updating, segal2015updating}, have been shown to engage neural systems associated with attention regulation and cognitive control. Previous neuroimaging studies have demonstrated altered dorsolateral prefrontal cortex activation in individuals with elevated anxiety during such tasks~\cite{balderston2017anxiety}. Despite these findings, comparatively limited work has investigated whether cognitively demanding emotional tasks elicit differential autonomic responses in socially anxious individuals, particularly using wearable physiological sensing modalities such as EDA.

In this study, we investigate whether wearable EDA signals can differentiate socially anxious (SA) and non-socially anxious (NSA) individuals during an emotionally salient cognitive task. To address this question, we conducted a controlled experimental study involving 50 participants (25 SA and 25 NSA), during which EDA signals were continuously acquired using a Shimmer wearable sensor across both resting-state and 2-back task conditions. We employed a standardized interval-based signal processing pipeline and complemented conventional statistical analysis with machine learning classification models to examine both univariate and multivariate autonomic signatures associated with social anxiety.

Our findings indicate that although both SA and NSA participants exhibit increased EDA responses during task engagement relative to baseline, no statistically significant group differences emerge during the cognitively demanding task condition. In contrast, machine learning models achieved above-chance discrimination between SA and NSA participants using resting-state EDA features (average AUC~=~0.73), while performance declined substantially during task conditions (average AUC~$\leq$~0.57). These findings suggest that cognitively demanding emotional tasks may induce a strong common autonomic response that obscures subtle baseline differences associated with social anxiety. More broadly, our results highlight the context-dependent nature of anxiety-related physiological signatures and indicate that socially relevant cognitive load, in the absence of explicit social-evaluative threat, may not sufficiently differentiate autonomic responses in socially anxious individuals.

Our key contributions are outlined below:
\begin{itemize}
\item We investigate the utility of wearable EDA signals as objective autonomic markers for differentiating socially anxious and non-socially anxious individuals during cognitively demanding emotional tasks.
\item We develop a standardized interval-based physiological analysis framework that enables comparison between resting-state and task-induced autonomic responses.

\item We combine conventional statistical analysis with multivariate machine learning models to examine both population-level and individual-level physiological signatures associated with social anxiety.

\item We demonstrate that resting-state EDA contains subtle multivariate signatures associated with SAD, while cognitively demanding emotional tasks induce shared autonomic responses that reduce group separability.

\item Our findings provide insights into the context-dependent behavior of wearable physiological biomarkers for anxiety assessment and contribute to the broader development of scalable digital mental health monitoring systems.
\end{itemize}

\begin{figure}
\begin{center}
\includegraphics[width=\linewidth]{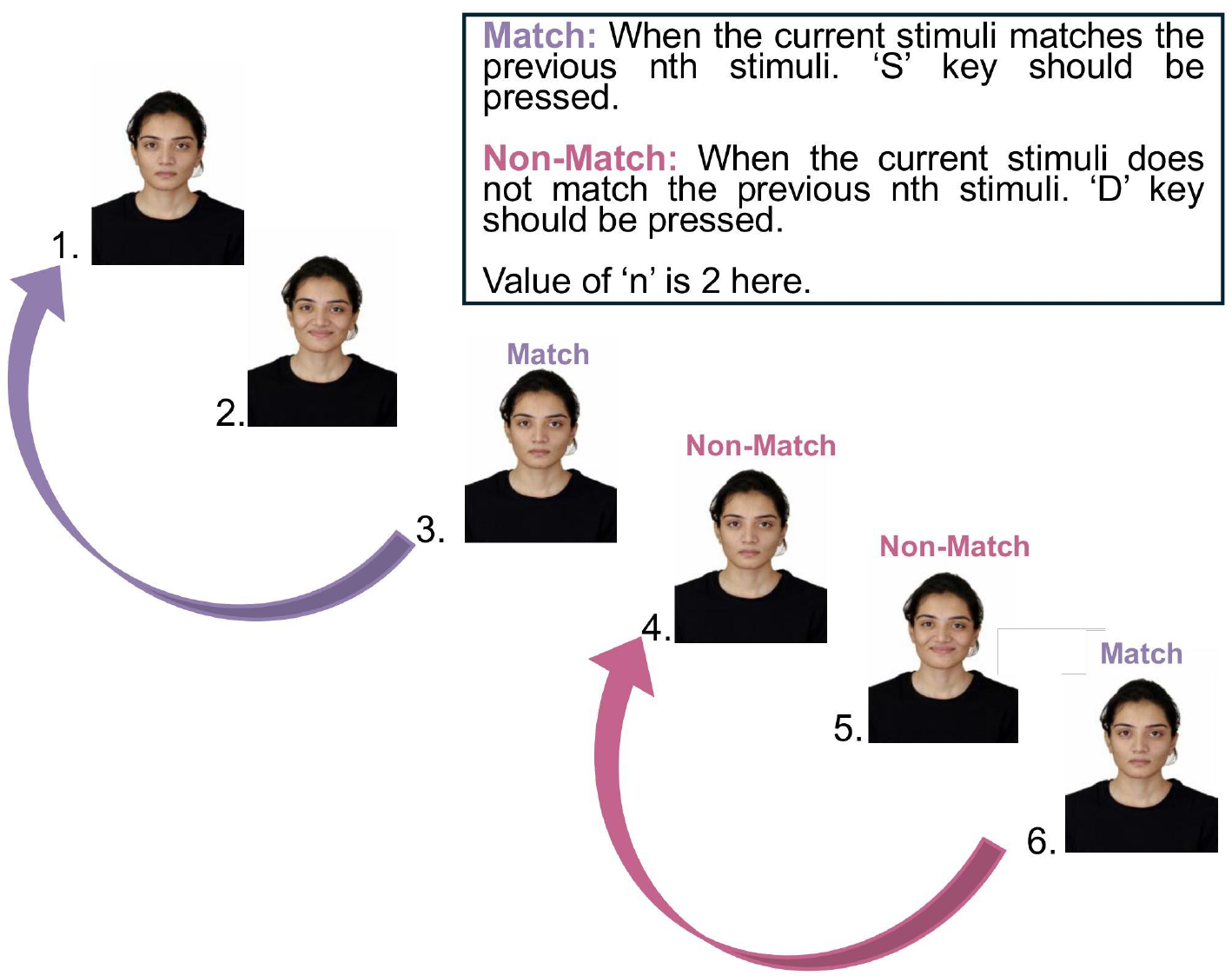} 
\end{center}
\caption{Explanation for the n-back task paradigm.}
\label{fig: n-back task}
\end{figure}

\section{Related Work}

\noindent{\it Wearable EDA Sensing for Anxiety Assessment:}
Wearable physiological sensing has emerged as a promising approach for objective mental health monitoring. Among various physiological modalities, electrodermal activity (EDA) is widely used as a non-invasive indicator of sympathetic autonomic nervous system activity and has shown potential as a digital biomarker for stress and anxiety assessment~\cite{boucsein2012electrodermal}. Prior studies have reported associations between elevated skin conductance and anxiety-related states across laboratory and real-world settings~\cite{mauss2003autonomic, wu2013impact, najafpour2017can}. Recent advances in wearable sensing and multimodal machine learning frameworks have further enabled scalable analysis of physiological responses during affective and cognitive tasks~\cite{bhatti2024clare}.

\noindent{\it EDA in Emotionally Salient Cognitive Tasks:}
EDA has been shown to reliably capture autonomic arousal during cognitively demanding and emotionally engaging tasks. Haapalainen et al.~\cite{haapalainen2010psycho} demonstrated increased EDA responses during complex visual and arithmetic tasks, while Nourbakhsh et al.~\cite{nourbakhsh2017detecting} observed elevated skin conductance during dual-task paradigms involving attentional and emotional demands. Similarly, Posada-Quintero et al.~\cite{posada2016power, perna2020personalized} reported that EDA is sensitive to emotionally stressful cognitive paradigms such as the Stroop task. These findings collectively support the use of EDA for monitoring autonomic responses during affective-cognitive interactions.

Despite growing interest in wearable anxiety sensing, relatively few studies have examined whether autonomic responses during emotionally salient cognitive tasks differ between socially anxious (SA) and non-socially anxious (NSA) individuals. Existing findings remain inconsistent across social stress paradigms, particularly in the absence of explicit social-evaluative threat. To address this gap, we investigate wearable EDA responses in SA and NSA participants during an emotional working memory task using a standardized physiological signal analysis pipeline and machine learning framework.

\section{Methodology}
\begin{figure}
\begin{center}
\includegraphics[width=0.5\linewidth]{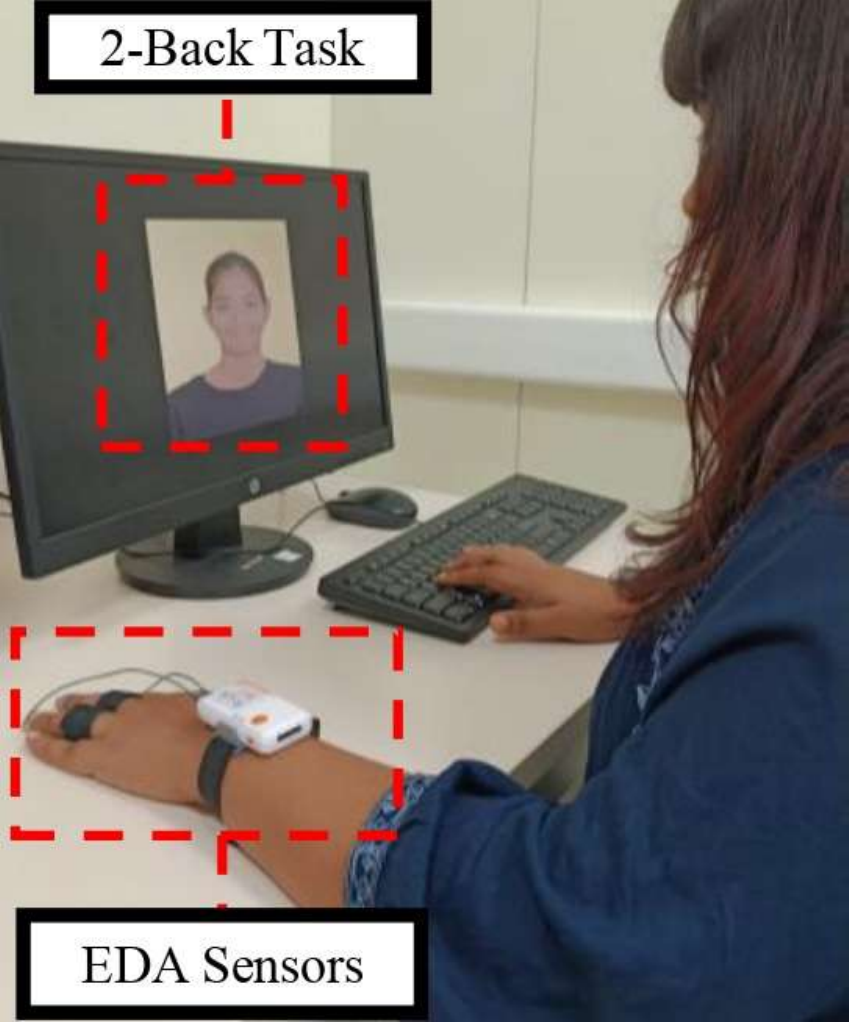} 
\end{center}
\caption{Experimental setup.} 
\label{fig:eda_sensor}

\end{figure}

\subsection{Participants}
Before proceeding with the study, ethics approval was obtained from the ethics committee of the IISER Bhopal Institute. The participant pool, within an age range of $18-30$, consists of two groups, both recruited through advertisement within the university campus. One group consists of sub-clinical population who met the diagnostic criteria for SAD as assessed by the \textit{Social Interaction Anxiety Scale (SIAS)} and the other group is of healthy individuals~\cite{brown1997validation}. SIAS is a 20-item self-report questionnaire that measures social interaction anxiety. Each item in the questionnaire is rated on a 5-point Likert scale, ranging from ``not at all characteristic of me" to ``extremely characteristic of me" making the total score range from 0 to 80. Participants scoring $\geq 43$ were classified as SA, while those scoring $\leq 33$ were classified as NSA, based on clinically accepted cut-offs. The sample consisted of equal numbers of SA and NSA individuals to facilitate comparative analysis. Demographic information, including age and gender, can be found in Table~\ref{tab:demographics}. To compensate them for their time, participants received either bonus points towards their course credit or a refreshment box containing snacks.

\begin{table}[ht]
\centering
\caption{Demographic details of the two experimental groups}
\begin{adjustbox}{max width=\textwidth}
\begin{tabular}{l c c c c c c c}
\toprule

\textbf{Group} & \textbf{N} & \textbf{Gender} & \textbf{Age} & \textbf{SIAS}  \\
 & & \textbf{(M/F)} & ($\overline{X} \pm \sigma$) & ($\overline{X} \pm \sigma$) \\
\midrule
SA   & 25 & 17/8 &$ 19.96 \pm 1.645$ & $51.76 \pm 8.001$  \\
NSA   & 25 & 16/9 & $19.72 \pm 1.54$2 & $18.56 \pm 6.621$ \\
\bottomrule
\end{tabular}
\end{adjustbox}
\label{tab:demographics}
\vspace{-2em}
\end{table}

 \begin{figure*}
\begin{center}
\includegraphics[width=0.9\linewidth]{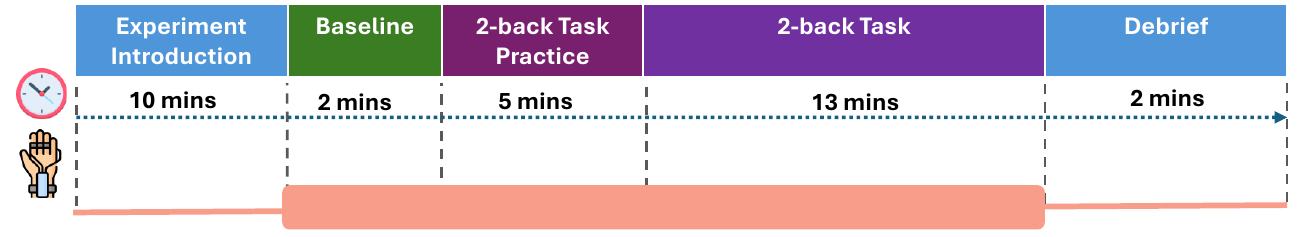} 
\end{center}
\caption{Timeline of a typical experimental session. The orange line corresponds to the EDA data acquisition timeline.}
\label{fig: master_diagram}
\end{figure*}

\subsection{Physiological Measure}
EDA data were collected using Shimmer3 GSR+ unit sensors\footnote{\url{https://www.shimmersensing.com/product/shimmer3-gsr-unit/}} specifically designed for clinical trials~\cite{burns2010shimmer}. Two EDA electrodes were placed near the base of the middle and ring finger on the participant's non-dominant hand to measure skin conductance. (as shown in Figure~\ref{fig:eda_sensor}). Participants' fingers were cleaned with alcohol wipes prior to sensor placement to remove residues and ensure optimal electrode-skin contact, and reducing signal noise in the EDA recordings. Data acquisition was performed at a sampling rate of 512 Hz during the time points as shown in Figure~\ref{fig: master_diagram}. After each study session, time-stamped EDA data were extracted from the Shimmer devices and securely saved on the study computer for further analysis. 


\subsection{Emotional 2-back Task}
Participants completed an emotional 2-back working memory task (Figure~\ref{fig: n-back task}) involving facial expressions (happy, neutral, and disgusted). For each trial, participants indicated whether the current facial expression matched the one presented two trials earlier. To minimize identity-related confounds, each block contained images of the same individual with varying facial expressions. Stimuli were obtained from the IIMI Emotional Face Database~\cite{iimi-dataset}. A total of 324 trials were presented across 12 randomized blocks. Please refer to~\cite{adyasha2025impact} for a detailed description of the task used.

\subsection{Procedure}
Each session began with informed consent  after the participant was briefed through a standard information sheet. This was followed by baseline EDA measurement for 2 minutes. They were then given a brief overview of the task with a practice session of 25 trials to get them accustomed to the task. The structure of the study is inspired by that measure EDA during a cognitive task~\cite{posada2016power}. Afterwards, they performed the main task that lasted up to 13 minutes. While EDA was obtained throughout the whole study, the data was later segmented to obtain EDA signals during baseline and during the 2-back task.

\subsection{Data Analysis Pipeline}

\subsubsection{Signal Preprocessing}

EDA signals were recorded at 512~Hz and downsampled to 32~Hz following anti-aliasing low-pass filtering. A fifth-order Butterworth low-pass filter (1~Hz cutoff) was subsequently applied to remove high-frequency noise and motion artifacts, following standard EDA preprocessing practices~\cite{moscato2023feasibility}. Signal decomposition into tonic and phasic components was performed using the convex optimization-based cvxEDA algorithm implemented in NeuroKit2~\cite{greco2015cvxeda, makowski2021neurokit2}.

\subsubsection{Feature Extraction}

Ten EDA features spanning tonic, phasic, spectral, sympathetic, and nonlinear domains were extracted from each analysis segment (baseline and task phases). These included tonic measures (EDA\_Tonic\_Mean, EDA\_Tonic\_SD), phasic measures (SCR\_Count, SCR\_Amplitude\_Mean), sympathetic indices (EDASymp, TVSymp, EDA\_Autocorrelation), spectral features (Spectral\_LF\_Power\_Norm, Spectral\_Entropy), and a nonlinear feature (Nonlinear\_DFA\_Alpha). Detailed feature descriptions are provided in Supplementary Table~A1. Tonic and phasic features were derived from cvxEDA decomposition, while spectral and nonlinear features were computed using Welch power spectral density estimation and detrended fluctuation analysis (DFA), respectively. Feature extraction was performed using NeuroKit2~\cite{makowski2021neurokit2} and SciPy.

\subsubsection{Temporal Analysis Design}

To examine temporal autonomic dynamics during task engagement, the first 6 minutes of the task were divided into three consecutive 2-minute phases: \textit{Early} (0--2 min), \textit{Middle}(2--4 min), and \textit{Late} (4--6 min). Each phase was compared against a duration-matched 2-minute baseline segment to avoid bias arising from unequal recording lengths~\cite{hossain2022comparison}. The same feature set was extracted independently from each phase for subsequent statistical analysis.

\begin{figure*}[!t]
    \includegraphics[width=1\linewidth]{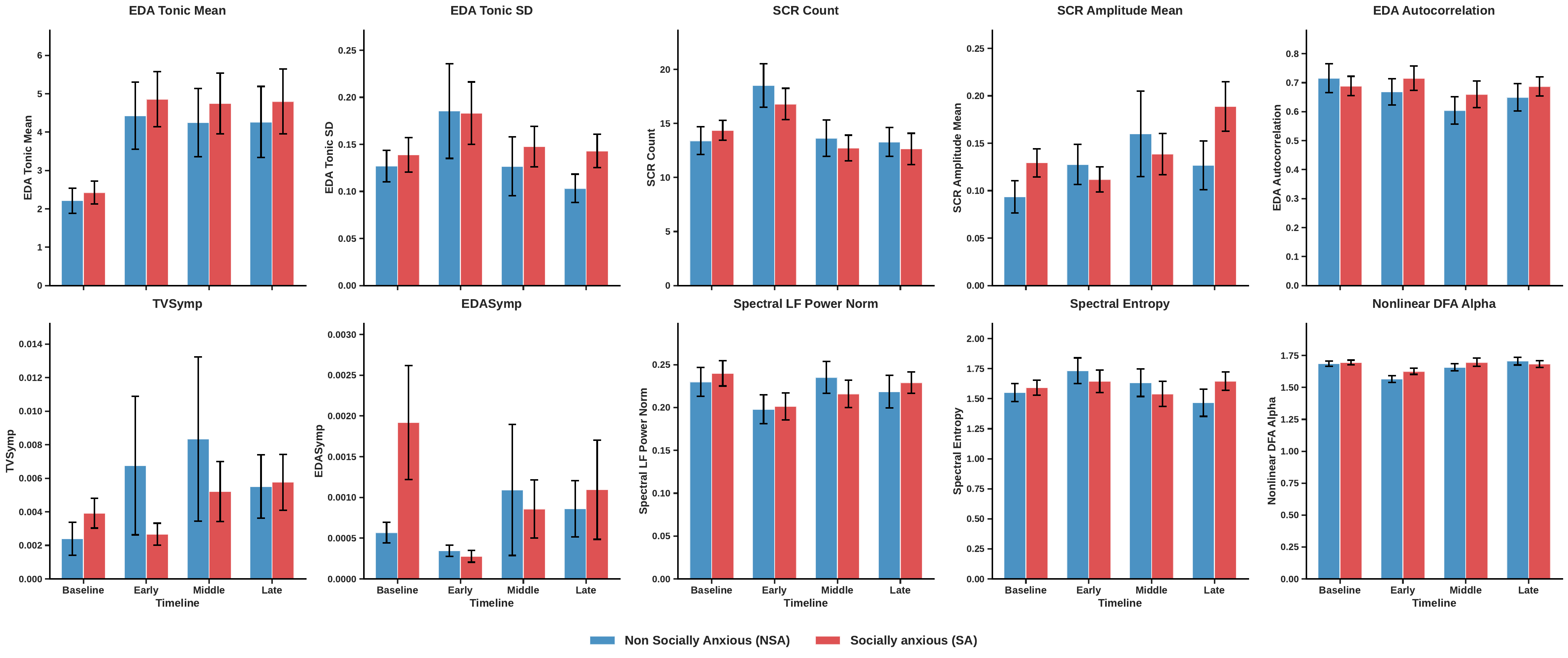}
   \caption{Mean EDA feature values across baseline and task phases for SA and NSA groups. Error bars represent standard error of the mean (SEM).}
\label{fig:bar_graph}
\vspace{-1.5em}
\end{figure*}

\subsection{Statistical Analysis}
For each task phase (Early, Middle, Late), a 2 (Condition: Baseline vs.\ Phase) $\times$ 2 (Group: SA vs.\ NSA) mixed ANOVA was performed, with Condition as a within-subject factor and Group as a between-subject factor. Normality and homogeneity assumptions were verified prior to analysis. Bonferroni-corrected post-hoc comparisons and simple effects analyses were conducted where appropriate. Effect sizes are reported using partial eta-squared ($\eta^2_p$) and Cohen's $d$.

Behavioral performance metrics from the 2-back task are reported in detail in~\cite{adyasha2025impact}; no significant group differences were observed in task accuracy or response time. All analyses were conducted in Python~3.10 using NeuroKit2, SciPy, and Pingouin. 
We release both our code and data publicly at \url{https://drive.google.com/drive/folders/1rbIuCZCxRPzQzIxO3BtZJuH5PDjGgau3?usp=sharing}

\begin{table*}[t]
\centering
\caption{ANOVA Results for EDA features at early phase of the task across experimental conditions and groups}
\begin{adjustbox}{max width=\textwidth}
\begin{tabular}{l c c c c c c c c c}
\toprule
\textbf{EDA Feature} 
& \multicolumn{3}{c}{\textbf{Condition Effect}} 
& \multicolumn{3}{c}{\textbf{Group Effect}} 
& \multicolumn{3}{c}{\textbf{Interaction (Condition $\times$ Group)}} \\
\cmidrule(lr){2-4} \cmidrule(lr){5-7} \cmidrule(lr){8-10}
& $F(1,48)$ & $p$ & $\eta^2_p$ 
& $F(1,48)$ & $p$ & $\eta^2_p$ 
& $F(1,48)$ & $p$ & $\eta^2_p$ \\
\midrule
EDA\_Tonic\_Mean (SCL)
& 36.698 & $< .001$ & 0.433 
& 0.176 & 0.677 & 0.004 
& 0.080 & 0.778 & 0.002 \\

EDA\_Tonic\_SD (SCL variability)
& 4.280 & .044 & 0.082 
& 0.015 & 0.902 & $< .001$ 
& 0.082 & 0.775 & 0.002 \\

SCR\_Count (NS-SCRs)
& 13.768 & $< .001$ & 0.223 
& 0.044 & 0.835 & 0.001 
& 1.730 & 0.195 & 0.035 \\

SCR\_Amplitude\_Mean 
& 1.306 & 0.259 & 0.026 
& 0.197 & 0.659 & 0.004 
& 12.634 & $< .001$ & 0.208 \\

EDA\_Autocorrelation 
& 0.082 & 0.776 & 0.002 
& 0.039 & 0.845 & 0.001 
& 1.103 & 0.299 & 0.022 \\

TVSymp 
& 0.612 & 0.438 & 0.013 
& 0.289 & 0.593 & 0.006 
& 2.002 & 0.164 & 0.040 \\

EDASymp 
& 7.327 & 0.009 & 0.132 
& 2.966 & 0.091 & 0.058 
& 4.239 & 0.045 & 0.081 \\

Spectral\_LF\_Power\_Norm 
& 5.523 & 0.023 & 0.103 
& 0.154 & 0.697 & 0.003 
& 0.046 & 0.830 & 0.001 \\

Spectral\_Entropy 
& 3.311 & 0.075 & 0.065 
& 0.053 & 0.819 & 0.001 
& 0.978 & 0.328 & 0.020 \\

Nonlinear\_DFA\_Alpha 
& 45.363 & $< .001$ & 0.486 
& 1.511 & 0.225 & 0.031 
& 3.427 & 0.070 & 0.067 \\

\bottomrule
\end{tabular}
\end{adjustbox}
\label{tab:eda_results_early}
\end{table*}

\begin{table*}[t]
\centering
\caption{ANOVA Results for EDA features at middle phase of the task across experimental conditions and groups}
\begin{adjustbox}{max width=\textwidth}
\begin{tabular}{l c c c c c c c c c}
\toprule
\textbf{EDA Feature} 
& \multicolumn{3}{c}{\textbf{Condition Effect}} 
& \multicolumn{3}{c}{\textbf{Group Effect}} 
& \multicolumn{3}{c}{\textbf{Interaction (Condition $\times$ Group)}} \\
\cmidrule(lr){2-4} \cmidrule(lr){5-7} \cmidrule(lr){8-10}
& $F(1,48)$ & $p$ & $\eta^2_p$ 
& $F(1,48)$ & $p$ & $\eta^2_p$ 
& $F(1,48)$ & $p$ & $\eta^2_p$ \\
\midrule
EDA\_Tonic\_Mean (SCL)
& 26.756 & $< .001$ & 0.358 
& 0.200 & 0.657 & 0.004 
& 0.115 & 0.736 & 0.002 \\

EDA\_Tonic\_SD (SCL Variability)
& 0.050 & 0.823 & 0.001 
& 0.411 & 0.524 & 0.008 
& 0.056 & 0.815 & 0.001 \\

SCR\_Count (NS-SCRs)
& 0.576 & 0.451 & 0.012 
& 0.000 & 0.990 & 0.000 
& 1.039 & 0.313 & 0.021 \\

SCR\_Amplitude\_Mean 
& 3.109 & 0.084 & 0.061 
& 0.050 & 0.825 & 0.001 
& 1.782 & 0.188 & 0.036 \\

EDA\_Autocorrelation 
& 3.143 & 0.083 & 0.061 
& 0.083 & 0.774 & 0.002 
& 1.103 & 0.299 & 0.022 \\

TVSymp 
& 2.204 & 0.144 & 0.044 
& 0.076 & 0.784 & 0.002 
& 0.916 & 0.343 & 0.019 \\

EDASymp 
& 0.252 & 0.618 & 0.005 
& 0.890 & 0.350 & 0.018 
& 2.180 & 0.146 & 0.043 \\

Spectral\_LF\_Power\_Norm 
& 0.313 & 0.578 & 0.006 
& 0.080 & 0.778 & 0.002 
& 0.764 & 0.386 & 0.016 \\

Spectral\_Entropy 
& 0.039 & 0.844 & 0.001 
& 0.068 & 0.796 & 0.001 
& 0.768 & 0.385 & 0.016 \\

Nonlinear\_DFA\_Alpha 
& 0.612 & 0.438 & 0.013 
& 0.559 & 0.459 & 0.012 
& 0.802 & 0.375 & 0.016 \\

\bottomrule
\end{tabular}
\end{adjustbox}
\label{tab:eda_middle_task_results}
\end{table*}

\begin{table*}[t]
\centering
\caption{ANOVA Results for EDA features at late phase of the task across experimental conditions and groups}
\begin{adjustbox}{max width=\textwidth}
\begin{tabular}{l c c c c c c c c c}
\toprule
\textbf{EDA Feature} 
& \multicolumn{3}{c}{\textbf{Condition Effect}} 
& \multicolumn{3}{c}{\textbf{Group Effect}} 
& \multicolumn{3}{c}{\textbf{Interaction (Condition $\times$ Group)}} \\
\cmidrule(lr){2-4} \cmidrule(lr){5-7} \cmidrule(lr){8-10}
& $F(1,48)$ & $p$ & $\eta^2_p$ 
& $F(1,48)$ & $p$ & $\eta^2_p$ 
& $F(1,48)$ & $p$ & $\eta^2_p$ \\
\midrule
EDA\_Tonic\_Mean (SCL)
& 23.631 & $< .001$ & 0.330 
& 0.208 & 0.650 & 0.004 
& 0.128 & 0.722 & 0.003 \\

EDA\_Tonic\_SD (SCL Variability)
& 0.701 & 0.407 & 0.014 
& 1.508 & 0.225 & 0.030 
& 1.442 & 0.236 & 0.029 \\

SCR\_Count (NS-SCRs)
& 0.746 & 0.392 & 0.015 
& 0.012 & 0.912 & 0.000 
& 0.564 & 0.456 & 0.012 \\

SCR\_Amplitude\_Mean 
& 13.790 & $< .001$ & 0.223 
& 3.135 & 0.083 & 0.061 
& 1.115 & 0.296 & 0.023 \\

EDA\_Autocorrelation 
& 1.020 & 0.318 & 0.021 
& 0.011 & 0.916 & 0.000 
& 0.947 & 0.335 & 0.019 \\

TVSymp 
& 6.575 & 0.014 & 0.120 
& 0.255 & 0.616 & 0.005 
& 0.434 & 0.513 & 0.009 \\

EDASymp 
& 0.405 & 0.527 & 0.008 
& 1.955 & 0.168 & 0.039 
& 1.772 & 0.189 & 0.036 \\

Spectral\_LF\_Power\_Norm 
& 0.576 & 0.452 & 0.012 
& 0.359 & 0.552 & 0.007 
& 0.001 & 0.980 & 0.000 \\

Spectral\_Entropy 
& 0.053 & 0.820 & 0.001 
& 1.222 & 0.275 & 0.025 
& 1.152 & 0.289 & 0.023 \\

Nonlinear\_DFA\_Alpha 
& 0.062 & 0.804 & 0.001 
& 0.053 & 0.819 & 0.001 
& 0.905 & 0.346 & 0.018 \\

\bottomrule
\end{tabular}
\end{adjustbox}
\label{tab:eda_late_task_results}
\end{table*}

\section{Results}

We present results organized around the three key findings that emerged from the temporal analysis. Full ANOVA results for all ten features across the three task phases are provided in Tables~\ref{tab:eda_results_early},~\ref{tab:eda_middle_task_results}, and~\ref{tab:eda_late_task_results}, and mean values are visualized in Figure~\ref{fig:bar_graph}. We first summarize the overall pattern, then detail each finding.

\subsection{Overview: Three Phases of Autonomic Response}

The temporal analysis revealed a clear trajectory across the 6-minute task period:
\begin{enumerate}
    \item \textbf{Broad initial activation} (Early phase): Six of ten features showed significant Condition effects, and two features showed significant Condition~$\times$~Group interactions (see Table~\ref{tab:eda_results_early}).
    \item \textbf{Rapid habituation} (Middle phase): Only one feature (EDA\_Tonic\_Mean) remained significantly elevated; all other effects, including all interactions, disappeared (see Table~\ref{tab:eda_middle_task_results}).
    \item \textbf{Late-phase qualitative shift} (Late phase): Tonic arousal persisted with reduced effect sizes, while two new effects emerged (SCR amplitude and TVSymp increased), reflecting a shift in the \textit{quality} rather than \textit{quantity} of sympathetic responses (see Table~\ref{tab:eda_late_task_results}).
\end{enumerate}

Critically, across all three phases and all ten features, \textit{no significant between-group main effects were observed}. The two interaction effects were confined to the Early phase and did not persist.

Complete post-hoc simple effects for all features are provided in supplementary Tables A.2, A.3, and A.4; all statistics reported in the text below can be verified against these tables.
 
\subsection{Finding 1: Task Onset Produces Robust, Group-Independent Autonomic Activation}

The 2-back task reliably elevated autonomic arousal across all participants, with the strongest effects at task onset. EDA\_Tonic\_Mean (SCL) increased from baseline ($M = 2.32 \pm 1.54~\mu S$) to the Early phase ($M = 4.64 \pm 3.92~\mu S$), $F(1,48) = 36.698$, $p < .001$, $\eta^2_p = .433$, $d = 0.865$. This elevation persisted into the Middle phase ($F(1,48) = 26.756$, $p < .001$, $\eta^2_p = .358$, $d = 0.738$) and Late phase ($F(1,48) = 23.631$, $p < .001$, $\eta^2_p = .330$, $d = 0.694$), with gradually declining effect sizes. Simple effects confirmed significant increases in both groups across all phases (all $p < .006$; see Tables A.2-- A.4 in supplementary), with no Group main effect or interaction in any phase (all $p > .650$).

Several other features showed significant Early-phase Condition effects that habituated by the Middle phase:
\begin{itemize}
    \item \textbf{SCR\_Count:} $F(1,48) = 13.768$, $p < .001$, $\eta^2_p = .223$, $d = 0.521$ (Baseline: $M = 13.88 \pm 5.51$; Early: $M = 17.66 \pm 8.65$). Simple effects: NSA $p = .004$, $d = 0.630$; SA $p = .058$, $d = 0.398$. By the Middle phase, this effect was absent ($p = .451$).
    
    \item \textbf{EDA\_Tonic\_SD:} $F(1,48) = 4.280$, $p = .044$, $\eta^2_p = .082$, $d = 0.295$. Neither within-group simple effect reached significance (NSA: $p = .180$; SA: $p = .100$). This effect should be interpreted cautiously.
    
    \item \textbf{Spectral\_LF\_Power\_Norm:} $F(1,48) = 5.523$, $p = .023$, $\eta^2_p = .103$, $d = -0.336$ (Baseline: $M = 0.235 \pm 0.078$; Early: $M = 0.200 \pm 0.080$). Recovered by the Middle phase ($p = .578$).
    
    \item \textbf{Nonlinear\_DFA\_Alpha:} The largest effect in the analysis: $F(1,48) = 45.363$, $p < .001$, $\eta^2_p = .486$, $d = -0.930$ (Baseline: $M = 1.691 \pm 0.098$; Early: $M = 1.596 \pm 0.125$). Both groups showed large decreases (NSA: $p < .001$, $d = -1.004$; SA: $p < .001$, $d = -0.943$). Recovered by the Middle phase ($p = .438$).
    \item \textbf{EDASymp:} $F(1,48) = 7.327$, $p = .009$, $\eta^2_p = .132$, $d = -0.371$ (Baseline: $M = 0.0012 \pm 0.0026$; Early: $M = 0.0003 \pm 0.0004$). This feature also showed a significant interaction, discussed in Finding~2.
\end{itemize}

Four features showed no significant Condition effects in the Early phase: SCR\_Amplitude\_Mean ($p = .259$), EDA\_Autocorrelation ($p = .776$), TVSymp ($p = .438$), and Spectral\_Entropy ($p = .075$). Of these, EDA\_Autocorrelation and Spectral\_Entropy remained non-significant across all phases (all $p > .075$).

\subsection{Finding 2: Two Transient Group Interactions in the Early Phase}

No between-group main effects were found in any phase. Two Condition~$\times$~Group interactions emerged in the Early phase only (see Table A.2 in supplementary for all simple effects).

\subsubsection{SCR Amplitude Mean}
The interaction was significant, $F(1,48) = 12.634$, $p < .001$, $\eta^2_p = .208$, while neither the Condition ($p = .259$) nor Group ($p = .659$) main effects were significant. Simple effects revealed a crossover pattern: NSA participants showed a significant increase from baseline ($t(24) = -3.230$, $p = .004$, $d = 0.646$), whereas SA participants showed a non-significant decrease ($t(24) = 1.756$, $p = .092$, $d = -0.351$). This interaction disappeared in the Middle phase ($p = .188$) and Late phase ($p = .296$).

In the Late phase, SCR\_Amplitude\_Mean showed a significant Condition main effect, $F(1,48) = 13.790$, $p < .001$, $\eta^2_p = .223$, with both groups increasing (NSA: $t(24) = -2.153$, $p = .042$, $d = 0.431$; SA: $t(24) = -3.031$, $p = .006$, $d = 0.606$). The Group effect approached significance ($p = .083$, $d = -0.501$) but the interaction was not significant ($p = .296$).

\subsubsection{EDASymp}
The interaction was significant, $F(1,48) = 4.239$, $p = .045$, $\eta^2_p = .081$. The SA group decreased significantly from baseline ($t(24) = 2.430$, $p = .023$, $d = -0.486$) while the NSA group did not change ($t(24) = 1.651$, $p = .112$, $d = -0.330$). At baseline, the SA group showed marginally higher EDASymp ($t(48) = -1.902$, $p = .063$, $d = -0.538$). However, this finding should be interpreted cautiously: the effect is relatively weak ($p = .045$), and the marginal baseline difference raises the possibility of a regression-to-the-mean artifact. This effect disappeared in the Middle ($p = .146$) and Late ($p = .189$) phases.

\subsection{Finding 3: Late-Phase Quantitative Shift in Sympathetic Response}

In the Late phase, while SCR frequency remained at baseline levels ($p = .392$), two features showed significant increases:

\begin{itemize}
    \item \textbf{SCR\_Amplitude\_Mean:} $F(1,48) = 13.790$, $p < .001$, $\eta^2_p = .223$ (Baseline: $M = 0.112 \pm 0.080$; Late: $M = 0.158 \pm 0.130$; $d = 0.525$). Both groups contributed: NSA $p = .042$, $d = 0.431$; SA $p = .006$, $d = 0.606$. No interaction ($p = .296$).
    
    \item \textbf{TVSymp:} $F(1,48) = 6.575$, $p = .014$, $\eta^2_p = .120$ (Baseline: $M = 0.003 \pm 0.005$; Late: $M = 0.006 \pm 0.009$; $d = 0.365$). Simple effects: NSA $p = .015$, $d = 0.523$; SA $p = .238$, $d = 0.242$. No interaction ($p = .513$).
\end{itemize}

This pattern--fewer but stronger SCRs alongside increased sympathetic variability--suggests a shift from broad initial reactivity to more focused sympathetic engagement during sustained task performance.

\section{Machine Learning Analysis}

Our statistical analysis established that group differences in EDA were absent or transient across task phases. To explore whether multivariate patterns in EDA data could detect subtle group differences that univariate tests may miss, we conducted classification analyses using both machine learning (ML) and deep learning (DL) approaches.

\subsection{Approach}
We used two complementary classification strategies. In the ML approach, we trained classifiers on the same 10 extracted EDA features used in the statistical analysis.
We selected four classifiers that represent diverse machine learning paradigms and have been widely validated in physiological signal classification tasks~\cite{schmidt2018introducing, greco2016arousal, sharma2012objective}: Logistic Regression (LR) as a baseline linear model known for interpretability and robustness with small sample sizes; Support Vector Machine (SVM) for its effectiveness in high-dimensional feature spaces and ability to capture nonlinear decision boundaries via kernel methods; Random Forest (RF) as an ensemble method that handles feature interactions and provides feature importance rankings; and Gradient Boosting (GB) for its sequential error-correction approach that often achieves state-of-the-art performance in structured data. This combination allows us to assess whether group differences are captured by linear versus nonlinear methods, and by single-model versus ensemble approaches. All classifiers were evaluated using 5-fold stratified cross-validation with standardized features to ensure fair comparison and account for our limited sample size (N=50).

In the DL approach, we trained a one-dimensional convolutional neural network (1D-CNN) directly on the raw EDA signal, following approaches successfully 
applied to physiological time-series classification~\cite{kiranyaz2015real, oh2018automated}. This tests whether the raw waveform contains discriminative patterns that feature extraction might miss. The raw EDA signal was downsampled from 512~Hz to 32~Hz, and 10-second sliding windows with 50\% overlap (step size = 5 seconds) were extracted from each 2-minute condition, yielding approximately 23 windows per participant per condition. Each window (320 samples) was independently
z-score normalized to account for inter-individual differences in absolute SCL. To prevent data leakage due to temporal overlap between adjacent windows, cross-validation splits were performed at the \textit{participant} level using grouped 5-fold cross-validation: all
windows from a given participant were assigned to the same fold, ensuring that the model was tested on entirely unseen individuals. 

The 1D-CNN architecture consisted of three convolutional layers (32, 64, 128 filters; kernel size = 5) with ReLU activation and max-pooling, followed by two fully connected layers (128, 2 units) with dropout (p = 0.3). The model was trained for 30 epochs using the Adam optimizer with binary cross-entropy loss. 
Cross-validation splits were performed at the participant level (i.e., all windows from a given participant appeared in the same fold) to prevent data leakage between training and test sets.

For both ML and DL, separate models were trained on baseline data and on each of the three task phases (Early, Middle, Late), using the same temporal segmentation as the statistical analysis. Classification performance was evaluated using two metrics: accuracy (Acc; proportion of correct classifications) and area under the receiver operating characteristic curve (AUC). AUC is reported as the primary metric because it provides a threshold-independent assessment of classifier discriminability across all possible decision thresholds~\cite{hanley1982meaning, bradley1997use}, making it more informative than accuracy alone, which depends on a single arbitrary threshold~\cite{lobo2008auc}. Additionally, AUC is the standard evaluation metric for binary classification in physiological and clinical prediction tasks~\cite{obuchowski2018receiver, mandrekar2010receiver}. An AUC of 0.5 indicates chance-level performance (no ability to distinguish SA from NSA), while an AUC of 1.0 indicates perfect classification.

\subsection{Results}
Classification results for all methods and conditions are presented in Table~\ref{tab:classification_results}.

\begin{table}[t]
\centering
\caption{Classification Performance (SA vs.\ NSA) Across Baseline and Task Phases}
\begin{adjustbox}{max width=\columnwidth}
\begin{tabular}{l l c c c c}
\toprule
\textbf{Method} & \textbf{Metric} & \textbf{Baseline} & \textbf{Early} & \textbf{Middle} & \textbf{Late}\\
\midrule
\multirow{2}{*}{LR}
& AUC & 0.67 & 0.56 & 0.42 & 0.62 \\
& Acc & 0.54 & 0.52 & 0.52 & 0.60 \\
\midrule
\multirow{2}{*}{SVM}
& AUC & 0.70 & 0.57 & 0.64 & 0.54 \\
& Acc & 0.60 & 0.52 & 0.58 & 0.52 \\
\midrule
\multirow{2}{*}{RF}
& AUC & \textbf{0.85} & 0.69 & 0.58 & 0.58 \\
& Acc & \textbf{0.82} & 0.60 & 0.56 & 0.56 \\
\midrule

\multirow{2}{*}{GB}
& AUC & 0.79 & 0.65 & 0.46 & 0.58 \\
& Acc & 0.66 & 0.58 & 0.48 & 0.56 \\
\midrule

\multirow{2}{*}{1D-CNN}
& AUC & 0.65 & 0.53 & 0.58 & 0.45 \\
& Acc & 0.62 & 0.52 & 0.54 & 0.46 \\
\bottomrule
\end{tabular}
\end{adjustbox}
\label{tab:classification_results}
\end{table}

\subsubsection{Baseline Classification}
Baseline EDA data yielded the best classification performance across all methods (see Table~\ref{tab:classification_results}). Among the ML classifiers, RF achieved the highest performance (AUC = 0.85, Accuracy = 0.82), followed by GB (AUC = 0.79) and SVM (AUC = 0.70). The 1D-CNN, which operated on raw EDA signals rather than extracted features, achieved an AUC of 0.65 and accuracy of 0.62. The average AUC across all five methods was 0.73, indicating above-chance discrimination between SA and NSA individuals based on resting-state EDA alone.

\subsubsection{Task Phase Classification}
Classification performance dropped substantially during all three task phases (see Table~\ref{tab:classification_results}). In the Early phase, the average AUC decreased from 0.73 (baseline) to 0.60. RF remained the best-performing ML classifier (AUC = 0.69), but this was considerably lower than its baseline performance. The 1D-CNN showed near-chance performance (AUC = 0.53).

Performance declined further in the Middle phase, with an average AUC of 0.54. Several classifiers fell at or below chance level (LR: AUC = 0.42; GB: AUC = 0.46; 1D-CNN: AUC = 0.58). This indicates that by the time participants had been performing the task for 2--4 minutes, the EDA signal no longer contained meaningful information for distinguishing between SA and NSA individuals.

In the Late phase, the average AUC was 0.55, showing minimal change from the Middle phase. No classifier achieved robust performance, and the 1D-CNN fell below chance (AUC = 0.45). The temporal pattern of declining classification performance across task phases mirrors the statistical finding that group-related interaction effects were confined to the Early phase and disappeared in later phases.

\subsubsection{Methodological Caveats}
Several caveats apply to the classification results. First, with $N = 50$ and 5-fold cross-validation, each test fold contains only 10 participants (5 per group); a single misclassification changes accuracy by 10\%, making performance estimates inherently noisy. Second, ensemble methods such as RF and GB are prone to overfitting with 10 features and 50 samples, even under cross-validation. Without nested cross-validation (with an inner loop for hyperparameter tuning and an outer loop for evaluation), the reported AUC values may be optimistically biased. Third, we do not report confidence intervals or permutation-based significance tests for the AUC estimates; the baseline RF result (AUC = 0.85) should therefore be interpreted as a \textit{promising signal} rather than a definitive finding, and replication with larger, independent samples is required before drawing conclusions about clinical utility. These caveats notwithstanding, the consistent \textit{temporal pattern}—best performance at baseline, declining to chance during the task—is robust across all five methods and reinforces the statistical findings.

\section{Discussion \& Conclusion}


This study investigated whether socially anxious (SA) and non-socially anxious (NSA) individuals exhibit differential autonomic responses during an emotionally salient working memory task using wearable electrodermal activity (EDA) sensing. Using a standardized physiological signal analysis pipeline and complementary machine learning analysis, we identified three primary findings.
\\
\noindent \textit{(i) The Task Increased Autonomic Arousal Equally in Both Groups:} The 2-back task successfully increased physiological arousal in all participants, confirming that the task was demanding enough to produce measurable sympathetic responses. However, across all EDA features and all  task phases, there were no significant differences between the SA and NSA groups. This means that social anxiety status did not affect how the body responded to this cognitive task. These findings suggest that emotionally salient cognitive load alone may be insufficient to elicit differential autonomic responses associated with SAD without explicit social-evaluative threat.
\\
\noindent \textit{(ii) Brief Group Differences Appeared Only at the Start:} Two small group differences appeared during the first two minutes of the task, but both disappeared quickly.

First, the SCR amplitude interaction ($\eta^2_p = .208$, $p < .001$) showed that NSA participants had stronger initial skin conductance responses when the task began, while SA participants showed slightly weaker responses compared to their own resting levels. This difference disappeared within two minutes as both groups settled into the task.

Second, the EDASymp interaction ($\eta^2_p = .081$, $p = .045$) showed that SA participants had slightly higher resting sympathetic activity, which dropped when the task started. However, this finding was weak and may simply reflect a statistical artifact (regression to the mean) rather than a real group difference. Future studies could address this by using baseline values as covariates or by collecting multiple baseline measurements.
\\
\noindent \textit{(iii) Late-Phase Shift: Fewer but Stronger Responses:} In the final phase, participants showed an interesting pattern: they produced fewer SCRs, but each response was stronger. This was accompanied by increased sympathetic variability (TVSymp). This suggests that as participants continued the task, their autonomic nervous system shifted from broad, frequent responses to fewer, more intense ones. Both groups showed this pattern, so it was not related to social anxiety.

\subsection{Machine Learning Results Support the Statistical Findings}
The machine learning findings converged closely with the statistical analysis. Resting-state EDA enabled moderate discrimination between SA and NSA groups (average AUC = 0.73), suggesting the presence of subtle multivariate autonomic differences at baseline. However, classification results declined to near-chance levels during task engagement across all models, indicating that the cognitively demanding task induced a shared autonomic response that reduced group separability.


However, once the task began, classification results dropped to near-chance levels (average AUC $\leq$ 0.57). All five classification methods-from simple logistic regression to a deep learning model-showed the same pattern: good performance at rest, poor performance during the task. This consistent decline across very different methods makes the finding reliable.

The baseline classification result (AUC = 0.85) should be treated cautiously because of the small sample size ($N = 50$) and the lack of advanced validation techniques. It is a promising signal, not a definitive finding, and needs to be confirmed with larger studies.

\subsection{Implications for Anxiety-Related Autonomic Responses}

Our findings are consistent with previous studies that found no EDA differences between SA and NSA individuals in non-evaluative settings~\cite{clark1992effects, mauss2003autonomic, pirinen2024associations, christian2023electrodermal, larrazabal2025understanding}. Even though our task used socially relevant stimuli (facial expressions), it did not trigger different physiological responses in socially anxious individuals because there was no social evaluation involved.

From  Attentional Control Theory (ACT)~\cite{eysenck2007anxiety}, which predicts that anxiety should disrupt physiological regulation under cognitive load, our results offer only partial support. The task did increase autonomic arousal, but both groups responded equally. This suggests that cognitive load alone is not enough-the evaluative aspect of a stressor may be necessary for anxiety-related physiological differences to appear.

Overall, our results show that EDA responses depend heavily on the context of the stressor. Trait anxiety alone does not determine physiological responses; the presence or absence of social-evaluative threat plays a critical role. These findings further suggest that wearable EDA-based biomarkers for social anxiety may be highly context dependent, with discriminative autonomic signatures emerging more reliably during socially evaluative stress rather than during cognitive load alone. This has important implications for the design of wearable digital mental health assessment systems.

\subsection{Limitations and Future Directions}

Our study has several limitations. First, participants were classified using the SIAS questionnaire, not clinical interviews. Some variability within groups may have reduced the differences between them. However, using self-report measures for group classification is standard practice in social anxiety research~\cite{shaukat2021detecting, larrazabal2025understanding, toner2025wearable, wang2023detecting, girondini2024decoupling}.

Second, the sample size of 50 participants is adequate for the statistical analyses but limits the reliability of the machine learning results. The promising baseline classification findings need to be confirmed with larger, independent samples.

Third, our study did not include a post-task recovery period. A recovery phase could reveal delayed group differences, as previous research suggests that socially anxious individuals may take longer to return to baseline after stress~\cite{mauss2003autonomic}.

Finally, future research should consider individual differences in coping strategies and emotion regulation, which can influence physiological responses even among people with high anxiety. Combining EDA with other measures such as heart rate variability or pupillometry could provide a more complete picture.

\subsection{Conclusion}

To answer our research question directly: SA and NSA individuals did \textit{not} show different EDA responses during the emotionally salient 2-back task. Both groups showed similar increases in autonomic arousal from baseline, and the only group-related differences were brief and disappeared within two minutes. Machine learning classification confirmed this-it could distinguish groups from resting-state EDA but not from task EDA, indicating that the task produced a strong, common physiological response that overshadowed any resting-state group differences. These results show that a cognitively demanding task with emotional content, but without social evaluation, does not produce different physiological responses in socially anxious individuals. Social-evaluative threat appears to be a necessary ingredient for the autonomic differences characteristic of social anxiety to emerge. 
 
\bibliographystyle{IEEEtran}
\bibliography{sample-base-2}

\begin{thebibliography}{10}
\providecommand{\url}[1]{#1}
\csname url@samestyle\endcsname
\providecommand{\newblock}{\relax}
\providecommand{\bibinfo}[2]{#2}
\providecommand{\BIBentrySTDinterwordspacing}{\spaceskip=0pt\relax}
\providecommand{\BIBentryALTinterwordstretchfactor}{4}
\providecommand{\BIBentryALTinterwordspacing}{\spaceskip=\fontdimen2\font plus
\BIBentryALTinterwordstretchfactor\fontdimen3\font minus \fontdimen4\font\relax}
\providecommand{\BIBforeignlanguage}[2]{{%
\expandafter\ifx\csname l@#1\endcsname\relax
\typeout{** WARNING: IEEEtran.bst: No hyphenation pattern has been}%
\typeout{** loaded for the language `#1'. Using the pattern for}%
\typeout{** the default language instead.}%
\else
\language=\csname l@#1\endcsname
\fi
#2}}
\providecommand{\BIBdecl}{\relax}
\BIBdecl

\bibitem{wells1995social}
A.~Wells, D.~M. Clark, P.~Salkovskis, J.~Ludgate, A.~Hackmann, and M.~Gelder, ``Social phobia: The role of in-situation safety behaviors in maintaining anxiety and negative beliefs,'' \emph{Behavior Therapy}, vol.~26, no.~1, pp. 153--161, 1995.

\bibitem{rapee1997cognitive}
R.~M. Rapee and R.~G. Heimberg, ``A cognitive-behavioral model of anxiety in social phobia,'' \emph{Behaviour research and therapy}, vol.~35, no.~8, pp. 741--756, 1997.

\bibitem{salari2024global}
N.~Salari, P.~Heidarian, M.~Hassanabadi, F.~Babajani, N.~Abdoli, M.~Aminian, and M.~Mohammadi, ``{Global Prevalence of Social Anxiety Disorder in Children, Adolescents and Youth: A Systematic Review and Meta-analysis},'' \emph{Journal of Prevention}, pp. 1--19, 2024.

\bibitem{kessler2005lifetime}
R.~C. Kessler, P.~Berglund, O.~Demler, R.~Jin, K.~R. Merikangas, and E.~E. Walters, ``Lifetime prevalence and age-of-onset distributions of dsm-iv disorders in the national comorbidity survey replication,'' \emph{Archives of general psychiatry}, vol.~62, no.~6, pp. 593--602, 2005.

\bibitem{stein2008social}
M.~B. Stein and D.~J. Stein, ``{Social anxiety disorder},'' \emph{The lancet}, vol. 371, no. 9618, pp. 1115--1125, 2008.

\bibitem{perna2020personalized}
G.~Perna, A.~Alciati, E.~Sangiorgio, D.~Caldirola, and C.~B. Nemeroff, ``{Personalized clinical approaches to anxiety disorders},'' \emph{Anxiety Disorders: Rethinking and Understanding Recent Discoveries}, pp. 489--521, 2020.

\bibitem{boucsein2012electrodermal}
W.~Boucsein, \emph{Electrodermal activity}.\hskip 1em plus 0.5em minus 0.4em\relax Springer Science \& Business Media, 2012.

\bibitem{sachdeva2022electrodermal}
P.~Sachdeva, S.~Ghosh, and J.~K. Sinha, ``Electrodermal activity (eda),'' in \emph{Encyclopedia of Sexual Psychology and Behavior}.\hskip 1em plus 0.5em minus 0.4em\relax Springer, 2022, pp. 1--6.

\bibitem{mauss2003autonomic}
I.~B. Mauss, F.~H. Wilhelm, and J.~J. Gross, ``Autonomic recovery and habituation in social anxiety,'' \emph{Psychophysiology}, vol.~40, no.~4, pp. 648--653, 2003.

\bibitem{wu2013impact}
T.~Wu, Y.~Luo, L.~S. Broster, R.~Gu, and Y.-j. Luo, ``{The impact of anxiety on social decision-making: Behavioral and electrodermal findings},'' \emph{Social neuroscience}, vol.~8, no.~1, pp. 11--21, 2013.

\bibitem{najafpour2017can}
E.~Najafpour, N.~Asl-Aminabadi, S.~Nuroloyuni, Z.~Jamali, and S.~Shirazi, ``Can galvanic skin conductance be used as an objective indicator of children’s anxiety in the dental setting?'' \emph{Journal of clinical and experimental dentistry}, vol.~9, no.~3, p. e377, 2017.

\bibitem{jensen1996electrodermal}
H.~H. JENSEN, N.~HASLE, and M.~BIRKET-SMITH, ``Electrodermal lability in anxiety disorders,'' \emph{Scandinavian journal of psychology}, vol.~37, no.~1, pp. 103--108, 1996.

\bibitem{gungor2020effect}
S.~Gungor, H.~Storm, J.~J. Bae, V.~Rotundo, and P.~J. Christos, ``The effect of emotional stressors on postoperative skin conductance indices: a prospective cohort pilot study,'' \emph{Revista Brasileira de Anestesiologia}, vol.~70, no.~4, pp. 325--332, 2020.

\bibitem{sahu2024wearable}
N.~K. Sahu, S.~Gupta, and H.~Lone, ``Wearable technology insights: Unveiling physiological responses during three different socially anxious activities,'' \emph{ACM Journal on Computing and Sustainable Societies}, vol.~2, no.~2, pp. 1--23, 2024.

\bibitem{moscovitch2010emotional}
D.~A. Moscovitch, M.~K. Suvak, and S.~G. Hofmann, ``Emotional response patterns during social threat in individuals with generalized social anxiety disorder and non-anxious controls,'' \emph{Journal of anxiety disorders}, vol.~24, no.~7, pp. 785--791, 2010.

\bibitem{eckman1997habituation}
P.~S. Eckman and G.~D. Shean, ``Habituation of cognitive and physiological arousal and social anxiety,'' \emph{Behaviour research and therapy}, vol.~35, no.~12, pp. 1113--1121, 1997.

\bibitem{fischer2021systematic}
S.~Fischer, F.~Haas, and J.~Strahler, ``A systematic review of thermosensation and thermoregulation in anxiety disorders,'' \emph{Frontiers in Physiology}, vol.~12, p. 784943, 2021.

\bibitem{white2018low}
E.~C. White and B.~M. Graham, ``{Low estradiol is linked to increased skin conductance, but not subjective anxiety or affect, in response to an impromptu speech task},'' \emph{Psychoneuroendocrinology}, vol.~98, pp. 30--38, 2018.

\bibitem{panayiotou2017psychophysiological}
G.~Panayiotou, M.~Karekla, D.~Georgiou, E.~Constantinou, and M.~Paraskeva-Siamata, ``{Psychophysiological and self-reported reactivity associated with social anxiety and public speaking fear symptoms: Effects of fear versus distress},'' \emph{Psychiatry research}, vol. 255, pp. 278--286, 2017.

\bibitem{heiser2009differentiating}
N.~A. Heiser, S.~M. Turner, D.~C. Beidel, and R.~Roberson-Nay, ``{Differentiating social phobia from shyness},'' \emph{Journal of anxiety disorders}, vol.~23, no.~4, pp. 469--476, 2009.

\bibitem{moran2016anxiety}
T.~P. Moran, ``Anxiety and working memory capacity: A meta-analysis and narrative review.'' \emph{Psychological bulletin}, vol. 142, no.~8, p. 831, 2016.

\bibitem{braver1997parametric}
T.~S. Braver, J.~D. Cohen, L.~E. Nystrom, J.~Jonides, E.~E. Smith, and D.~C. Noll, ``{A parametric study of prefrontal cortex involvement in human working memory},'' \emph{Neuroimage}, vol.~5, no.~1, pp. 49--62, 1997.

\bibitem{smith1997working}
E.~E. Smith and J.~Jonides, ``{Working memory: A view from neuroimaging},'' \emph{Cognitive psychology}, vol.~33, no.~1, pp. 5--42, 1997.

\bibitem{levens2010updating}
S.~M. Levens and I.~H. Gotlib, ``{Updating positive and negative stimuli in working memory in depression.}'' \emph{Journal of Experimental Psychology: General}, vol. 139, no.~4, p. 654, 2010.

\bibitem{segal2015updating}
A.~Segal, Y.~Kessler, and G.~E. Anholt, ``{Updating the emotional content of working memory in social anxiety},'' \emph{Journal of behavior therapy and experimental psychiatry}, vol.~48, pp. 110--117, 2015.

\bibitem{balderston2017anxiety}
N.~L. Balderston, K.~E. Vytal, K.~O'Connell, S.~Torrisi, A.~Letkiewicz, M.~Ernst, and C.~Grillon, ``{Anxiety patients show reduced working memory related dlPFC activation during safety and threat},'' \emph{Depression and anxiety}, vol.~34, no.~1, pp. 25--36, 2017.

\bibitem{bhatti2024clare}
A.~Bhatti, P.~Angkan, B.~Behinaein, Z.~Mahmud, D.~Rodenburg, H.~Braund, P.~J. Mclellan, A.~Ruberto, G.~Harrison, D.~Wilson \emph{et~al.}, ``{CLARE: Cognitive Load Assessment in REaltime with Multimodal Data},'' \emph{arXiv preprint arXiv:2404.17098}, 2024.

\bibitem{haapalainen2010psycho}
E.~Haapalainen, S.~Kim, J.~F. Forlizzi, and A.~K. Dey, ``Psycho-physiological measures for assessing cognitive load,'' in \emph{Proceedings of the 12th ACM international conference on Ubiquitous computing}, 2010, pp. 301--310.

\bibitem{nourbakhsh2017detecting}
N.~Nourbakhsh, F.~Chen, Y.~Wang, and R.~A. Calvo, ``Detecting users’ cognitive load by galvanic skin response with affective interference,'' \emph{ACM Transactions on Interactive Intelligent Systems (TiiS)}, vol.~7, no.~3, pp. 1--20, 2017.

\bibitem{posada2016power}
H.~F. Posada-Quintero, J.~P. Florian, A.~D. Orjuela-Ca{\~n}{\'o}n, T.~Aljama-Corrales, S.~Charleston-Villalobos, and K.~H. Chon, ``Power spectral density analysis of electrodermal activity for sympathetic function assessment,'' \emph{Annals of biomedical engineering}, vol.~44, pp. 3124--3135, 2016.

\bibitem{brown1997validation}
E.~J. Brown, J.~Turovsky, R.~G. Heimberg, H.~R. Juster, T.~A. Brown, and D.~H. Barlow, ``{Validation of the Social Interaction Anxiety Scale and the Social Phobia Scale across the anxiety disorders.}'' \emph{Psychological assessment}, vol.~9, no.~1, p.~21, 1997.

\bibitem{burns2010shimmer}
A.~Burns, B.~R. Greene, M.~J. McGrath, T.~J. O'Shea, B.~Kuris, S.~M. Ayer, F.~Stroiescu, and V.~Cionca, ``{SHIMMER™--A wireless sensor platform for noninvasive biomedical research},'' \emph{IEEE Sensors Journal}, vol.~10, no.~9, pp. 1527--1534, 2010.

\bibitem{iimi-dataset}
\BIBentryALTinterwordspacing
S.~Tewari, S.~Mehta, and N.~Srinivasan, ``{IIMI Emotional Face Database},'' 2023, accessed: 2026-04-20. [Online]. Available: \url{osf.io/f7zbv}
\BIBentrySTDinterwordspacing

\bibitem{adyasha2025impact}
A.~Adyasha, A.~S. Shelke, N.~K. Sahu, and H.~R. Lone, ``The impact of acute social stress on working memory updating in social anxiety disorder,'' in \emph{Proceedings of the Annual Meeting of the Cognitive Science Society}, vol.~47, 2025.

\bibitem{moscato2023feasibility}
S.~Moscato, S.~Orlandi, G.~Battaglia, F.~Di~Gregorio, G.~Lullini, S.~Pozzi, L.~Sabattini, L.~Chiari, and F.~La~Porta, ``Feasibility of a diagnostic differentiation tool for nociceptive and neuropathic pain in a neurorehabilitation population using physiological data from wearable sensors,'' in \emph{2023 IEEE EMBS Special Topic Conference on Data Science and Engineering in Healthcare, Medicine and Biology}.\hskip 1em plus 0.5em minus 0.4em\relax IEEE, 2023, pp. 45--46.

\bibitem{greco2015cvxeda}
A.~Greco, G.~Valenza, A.~Lanata, E.~P. Scilingo, and L.~Citi, ``cvxeda: A convex optimization approach to electrodermal activity processing,'' \emph{IEEE transactions on biomedical engineering}, vol.~63, no.~4, pp. 797--804, 2015.

\bibitem{makowski2021neurokit2}
D.~Makowski, T.~Pham, Z.~J. Lau, J.~C. Brammer, F.~Lespinasse, H.~Pham, C.~Sch{\"o}lzel, and S.~A. Chen, ``Neurokit2: A python toolbox for neurophysiological signal processing,'' \emph{Behavior research methods}, pp. 1--8, 2021.

\bibitem{hossain2022comparison}
M.-B. Hossain, Y.~Kong, H.~F. Posada-Quintero, and K.~H. Chon, ``Comparison of electrodermal activity from multiple body locations based on standard eda indices’ quality and robustness against motion artifact,'' \emph{Sensors}, vol.~22, no.~9, p. 3177, 2022.

\bibitem{schmidt2018introducing}
P.~Schmidt, A.~Reiss, R.~Duerichen, C.~Marberger, and K.~Van~Laerhoven, ``Introducing wesad, a multimodal dataset for wearable stress and affect detection,'' in \emph{Proceedings of the 20th ACM international conference on multimodal interaction}, 2018, pp. 400--408.

\bibitem{greco2016arousal}
A.~Greco, G.~Valenza, L.~Citi, and E.~P. Scilingo, ``Arousal and valence recognition of affective sounds based on electrodermal activity,'' \emph{IEEE Sensors Journal}, vol.~17, no.~3, pp. 716--725, 2016.

\bibitem{sharma2012objective}
N.~Sharma and T.~Gedeon, ``Objective measures, sensors and computational techniques for stress recognition and classification: A survey,'' \emph{Computer methods and programs in biomedicine}, vol. 108, no.~3, pp. 1287--1301, 2012.

\bibitem{kiranyaz2015real}
S.~Kiranyaz, T.~Ince, and M.~Gabbouj, ``Real-time patient-specific ecg classification by 1-d convolutional neural networks,'' \emph{IEEE transactions on biomedical engineering}, vol.~63, no.~3, pp. 664--675, 2015.

\bibitem{oh2018automated}
S.~L. Oh, E.~Y. Ng, R.~San~Tan, and U.~R. Acharya, ``Automated diagnosis of arrhythmia using combination of cnn and lstm techniques with variable length heart beats,'' \emph{Computers in biology and medicine}, vol. 102, pp. 278--287, 2018.

\bibitem{hanley1982meaning}
J.~A. Hanley and B.~J. McNeil, ``The meaning and use of the area under a receiver operating characteristic (roc) curve.'' \emph{Radiology}, vol. 143, no.~1, pp. 29--36, 1982.

\bibitem{bradley1997use}
A.~P. Bradley, ``The use of the area under the roc curve in the evaluation of machine learning algorithms,'' \emph{Pattern recognition}, vol.~30, no.~7, pp. 1145--1159, 1997.

\bibitem{lobo2008auc}
J.~M. Lobo, A.~Jim{\'e}nez-Valverde, and R.~Real, ``Auc: a misleading measure of the performance of predictive distribution models,'' \emph{Global ecology and Biogeography}, vol.~17, no.~2, pp. 145--151, 2008.

\bibitem{obuchowski2018receiver}
N.~A. Obuchowski and J.~A. Bullen, ``Receiver operating characteristic (roc) curves: review of methods with applications in diagnostic medicine,'' \emph{Physics in Medicine \& Biology}, vol.~63, no.~7, p. 07TR01, 2018.

\bibitem{mandrekar2010receiver}
J.~N. Mandrekar, ``Receiver operating characteristic curve in diagnostic test assessment,'' \emph{Journal of thoracic oncology}, vol.~5, no.~9, pp. 1315--1316, 2010.

\bibitem{clark1992effects}
B.~M. Clark, D.~A. Siddle, and N.~W. Bond, ``Effects of social anxiety and facial expression on habituation of the electrodermal orienting response,'' \emph{Biological Psychology}, vol.~33, no. 2-3, pp. 211--223, 1992.

\bibitem{pirinen2024associations}
V.~Pirinen, K.~Eggers, K.~Dindar, T.~Helminen, A.~Kotila, S.~Kuusikko-Gauffin, L.~M{\"a}kinen, H.~Ebeling, T.~Hurtig, M.~M{\"a}ntymaa \emph{et~al.}, ``Associations between social anxiety, physiological reactivity, and speech disfluencies in autistic young adults and controls,'' \emph{Journal of Communication Disorders}, vol. 109, p. 106425, 2024.

\bibitem{christian2023electrodermal}
C.~Christian, E.~Cash, D.~A. Cohen, C.~M. Trombley, and C.~A. Levinson, ``Electrodermal activity and heart rate variability during exposure fear scripts predict trait-level and momentary social anxiety and eating-disorder symptoms in an analogue sample,'' \emph{Clinical Psychological Science}, vol.~11, no.~1, pp. 134--148, 2023.

\bibitem{larrazabal2025understanding}
M.~A. Larrazabal, Z.~Wang, M.~Rucker, E.~R. Toner, M.~Boukhechba, B.~A. Teachman, and L.~E. Barnes, ``Understanding state social anxiety in virtual social interactions using multimodal wearable sensing indicators,'' \emph{arXiv preprint arXiv:2503.15637}, 2025.

\bibitem{eysenck2007anxiety}
M.~W. Eysenck, N.~Derakshan, R.~Santos, and M.~G. Calvo, ``Anxiety and cognitive performance: attentional control theory.'' \emph{Emotion}, vol.~7, no.~2, p. 336, 2007.

\bibitem{shaukat2021detecting}
R.~Shaukat-Jali, N.~van Zalk, D.~E. Boyle \emph{et~al.}, ``Detecting subclinical social anxiety using physiological data from a wrist-worn wearable: small-scale feasibility study,'' \emph{JMIR Formative Research}, vol.~5, no.~10, p. e32656, 2021.

\bibitem{toner2025wearable}
E.~R. Toner, M.~Rucker, Z.~Wang, M.~A. Larrazabal, L.~Cai, D.~Datta, H.~Lone, M.~Boukhechba, B.~A. Teachman, and L.~E. Barnes, ``Wearable sensor-based multimodal physiological responses of socially anxious individuals in social contexts on zoom,'' \emph{IEEE Transactions on Affective Computing}, 2025.

\bibitem{wang2023detecting}
Z.~Wang, M.~A. Larrazabal, M.~Rucker, E.~R. Toner, K.~E. Daniel, S.~Kumar, M.~Boukhechba, B.~A. Teachman, and L.~E. Barnes, ``Detecting social contexts from mobile sensing indicators in virtual interactions with socially anxious individuals,'' \emph{Proceedings of the ACM on interactive, mobile, wearable and ubiquitous technologies}, vol.~7, no.~3, pp. 1--26, 2023.

\bibitem{girondini2024decoupling}
M.~Girondini, I.~Frigione, M.~Marra, M.~Stefanova, M.~Pillan, A.~Maravita, and A.~Gallace, ``Decoupling the role of verbal and non-verbal audience behavior on public speaking anxiety in virtual reality using behavioral and psychological measures,'' \emph{Frontiers in Virtual Reality}, vol.~5, p. 1347102, 2024.

\end{thebibliography}

\end{document}